\documentclass[%
 preprint,
 superscriptaddress,
 amsmath,amssymb,
 aps, prapplied,
]{revtex4-2}

\usepackage{graphicx}
\usepackage{dcolumn}
\usepackage{bm}
\usepackage{siunitx}
\DeclareSIUnit{\sample}{Sa} 
\usepackage{hyperref}

\begin{document}

\title{\textbf{High-Speed NV Ensemble Magnetic Field Imaging via Laser Raster Scanning}}

\author{Luca Troise}
\email{lutro@dtu.dk}
\affiliation{Center for Macroscopic Quantum States (bigQ), Department of Physics, Technical University of Denmark, DK-2800 Kongens Lyngby, Denmark}

\author{Nikolaj W. Hansen}
\affiliation{Department of Neuroscience, University of Copenhagen, 2200 Copenhagen, Denmark}

\author{Marvin Holten}
\affiliation{DiaSense ApS, 2800 Kongens Lyngby, Denmark}

\author{Dhiren Kara}
\affiliation{Center for Macroscopic Quantum States (bigQ), Department of Physics, Technical University of Denmark, DK-2800 Kongens Lyngby, Denmark}

\author{Jean-Fran\c{c}ois Perrier}
\affiliation{Department of Neuroscience, University of Copenhagen, 2200 Copenhagen, Denmark}

\author{Ulrik L. Andersen}
\affiliation{Center for Macroscopic Quantum States (bigQ), Department of Physics, Technical University of Denmark, DK-2800 Kongens Lyngby, Denmark}

\author{Alexander Huck}
\email{alhu@dtu.dk}
\affiliation{Center for Macroscopic Quantum States (bigQ), Department of Physics, Technical University of Denmark, DK-2800 Kongens Lyngby, Denmark}

\date{\today}

\begin{abstract}
We present a technique that uses an ensemble of nitrogen-vacancy (NV) centers in diamond to image magnetic fields with high spatio-temporal resolution and sensitivity. A focused laser beam is raster-scanned using an acousto-optic deflector (AOD) and NV center fluorescence is read out with a single photodetector, enabling low-noise detection with high dynamic range. The method operates in a previously unexplored regime, quasi-continuous wave optically detected magnetic resonance (qCW-ODMR). In this regime, NV centers experience short optical pump pulses for spin readout and repolarization—analogous to pulsed ODMR—while the microwave field continuously drives the spin transitions.
We systematically characterize this regime and show that the spin response is governed by a tunable interplay between coherent evolution and relaxation, determined by the temporal spacing between pump laser pulses. Notably, the technique does not require precise microwave pulse control, thus simplifying experimental implementation. To demonstrate its capabilities, we image time-varying magnetic fields from a microwire with sub-millisecond temporal resolution. This approach enables flexible spatial sampling and, with our diamond, achieves \SI{}{\nano\tesla\per\sqrt\hertz}-level per-pixel sensitivity, making it well-suited for detecting weak, dynamic magnetic fields in biological and other complex systems.
\end{abstract}

\maketitle


\section{\label{sec:level1}Introduction}
Wide-field magnetic imaging utilizing paramagnetic color centers in the solid state is an emerging platform with a broad range of applications, ranging from materials science to life sciences. The nitrogen-vacancy (NV) centers in diamond are a particularly powerful system due to their long electron spin coherence time, operation under ambient conditions and stability~\cite{doherty2013nitrogen, degen2017quantum}. Ensembles of NV color centers with optimized concentration and isotopically engineered diamond host materials offer exceptional levels of sensitivity combined with microscale spatial resolution and large detection bandwidth, tunable with spin-control sequences. Several recent advances have demonstrated their utility for both wide-field imaging and quantum-enhanced performance~\cite{barry2020sensitivity, bauch2018ultralong}. Wide-field NV imaging has been applied to the study of current flow and electronic transport in devices ranging from integrated circuits to two-dimensional materials~\cite{nowodzinski2015nitrogen, tetienne2017quantum, webb2022high, horsley2018microwave}, the visualization of magnetic textures, including vortices in superconductors~\cite{schlussel2018wide, lillie2020laser}, and for geological investigations~\cite{glenn2017micrometer, fu2020high}. In biological contexts, NV-based wide-field methods have also been employed to image magnetotactic organisms and single cells with internalized nanoparticles~\cite{le2013optical, glenn2015single}. Importantly, this wide-field modality also enables the detection of magnetic fields generated by electrical activity in neurons or muscle tissue~\cite{barry2016optical, webb2021detection, troise2022vitro, hansen2023microscopic}.

Despite these successes, imaging biomagnetic processes remains a challenging task. The underlying magnetic signals are typically only tens to hundreds of picotesla in amplitude and vary on millisecond timescales, demanding detection bandwidths in the kilohertz range or higher. Capturing these low-amplitude, fast-varying fields with spatial resolution imposes strict requirements on both sensitivity and recording bandwidth~\cite{webb2020optimization}. In NV-based magnetic field imaging, a challenge arises from the fact that the magnetic information is encoded as a small change in fluorescence intensity superimposed on a large, static background. Conventional CMOS and CCD cameras, commonly used for wide-field readout, further exacerbate this issue by imposing trade-offs between temporal resolution, digitization depth, and noise levels, ultimately limiting sensitivity by technical rather than physical constraints~\cite{wojciechowski2018contributed}.

An alternative to camera-based wide-field magnetic imaging is to raster-scan the excitation laser beam across the layer containing the ensemble of NV centers and collect the resulting fluorescence with a photodetector. This approach encodes the spatial information serially in time rather than in parallel on a camera chip. In this case, the achievable detection bandwidth is set by the steering hardware and, ultimately, by the spin dynamics of the NV centers. Various beam-steering methods can be employed, such as galvanometric mirrors~\cite{keppler2022dynamic} or electro-optic deflectors. In particular, acousto-optic deflectors (AODs) are emerging as a powerful solution, offering beam steering with sub-\SI{}{\micro\second} temporal control. Well established in cold-atom experiments~\cite{endres2016atom, kaufman2021quantum}, AODs are now finding applications in a range of color-center–based techniques, including single-spin control~\cite{cambria2024scalable}, nanoscale NMR spectroscopy~\cite{leibold2024time}, and confocal-based imaging~\cite{ang2025design}.

\begin{figure}[t]  
    \centering
    \includegraphics[width=0.7\linewidth]{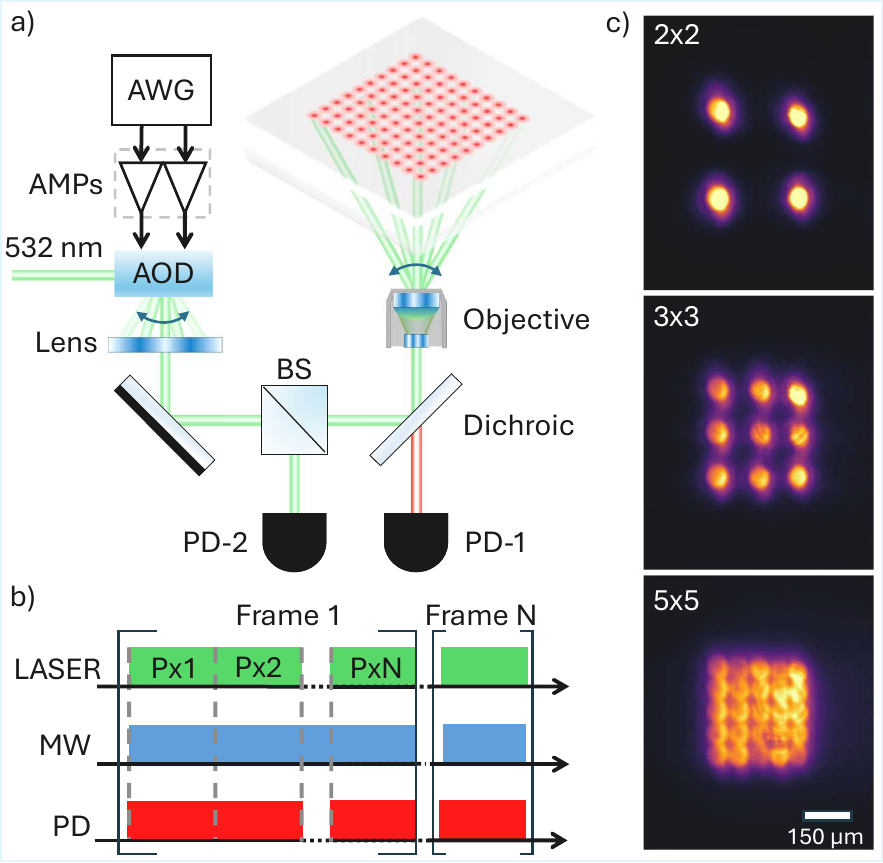}
    \caption{
(a) Schematic of the scanning-based NV magnetometry setup. The beam of a 532~nm laser is raster-scanned across the layer of NV centers using a 2D AOD, with fluorescence detected by a PIN photodetector. A second photodetector provides a reference for common-mode rejection and technical noise suppression.  
(b) qCW-ODMR diagram: laser scanning (green), continuous microwave excitation (blue), and photodetector readout (red) across successive pixels and frames.
(c) Fluorescence images of the diamond captured with a camera from above for different scan patterns as indicated in the images.  
}
    \label{fig:figure1}  
\end{figure}
In this work, we implement a novel laser-scanning wide-field imaging scheme based on a two-axis AOD system. A focused excitation beam is raster-scanned across the ensemble of NV centers embedded in a \SI{25}{\micro\meter}-thick layer, while the resulting fluorescence is recorded with a PIN photodiode (Fig.~\ref{fig:figure1}(a)). Among its advantages, first, digitizing the detector output with a 16-bit data acquisition system overcomes the bit-depth constraints of typical CMOS/CCD cameras \cite{manin2018performance}, enabling high dynamic-range measurement of fluorescence changes imposed on a large background noise. Second, the high bandwidth of the scanning device allows for high temporal resolution, with the pixel dwell time limited primarily by the optically induced spin-repolarization time of the NV centers. This enables resolving fluorescence transients with sub-microsecond precision — well beyond the capabilities of conventional cameras. Third, the use of a reference photodetector allows for the implementation of common-mode rejection (CMR) in real time, significantly reducing the noise floor in the presence of technical laser noise. Together, these features provide flexibility to optimize the wide-field magnetic imaging setup for sensitivity, spatial resolution, field of view, and temporal resolution, parameters that may be tuned depending on the specific requirements of the magnetic imaging task.

Building on the excitation beam scanning approach, we implement a hybrid magnetic field imaging protocol as illustrated by the sequence shown in Fig.~\ref{fig:figure1}b. The NV center electron spins are continuously driven by the applied microwave field while being read out and repolarized with short pulses of the excitation laser. Since the excitation beam is scanned across the sample, each voxel experiences the excitation as a brief pulse with a duration set by the pixel dwell time. We refer to this regime as quasi-continuous wave ODMR (qCW-ODMR). Because the excitation is spatially serialized, increasing the number of pixels per frame increases the spacing between consecutive laser excitation pulses at a given location, yielding a modified electron spin evolution and thus impacting sensitivity. To quantify these effects, we vary the pulse spacing and characterize the sensor performance in terms of ODMR contrast, linewidth, and noise floor.
Finally, to validate the capability of our technique, we image the time-varying magnetic field induced by a current pulse on a \SI{20}{\micro\meter} copper wire, resolving spatio-temporal fluctuations of the magnetic field in real time.

Our work establishes AOD-based scanning of the pump laser beam in combination with photodetection of fluorescence with a PIN diode as a viable alternative to camera-based NV imaging, unlocking new possibilities for high-speed, high-sensitivity NV-based wide-field magnetic imaging.

\section{\label{sec:level2}Methods}

\subsection{Optical Setup and NV center Excitation}

Our setup is powered by a \SI{532}{\nano\meter} continuous-wave laser delivering around \SI{300}{\milli\watt} of power measured in front of a 100x/0.7 NA objective (Mitutoyo) with \num{12}~mm working distance and focused to a spot diameter of approximately \SI{10}{\micro\meter} in the diamond. The diamond sample is a \(2\times2\times0.5\)~\SI{}{\milli\meter\cubed} single-crystal diamond with a (100) surface orientation, with a \SI{25}{\micro\meter}-thick layer of $^{12}C$-enriched diamond doped with approximately \num{1}~ppm $^{14}$NV centers and mounted on a planar microwave resonator.

Before the microscope objective, the laser passes through a two-axis AOD (Isomet) that scans the focused laser beam in the plane of the NV-doped layer in the diamond. In our configuration, the AOD drive signal is centered at \SI{100}{\mega\hertz} with a \SI{\pm25}{\mega\hertz} bandwidth and provided by an arbitrary waveform generator (AWG, Tektronix) with a sampling rate of \SI{600}{\mega\sample\per\second}. With the frequency pattern delivered by the AWG, we effectively control the scanning pattern (see Fig.~\ref{fig:figure1}c for examples), the pixel dwell time $\tau_d$, the number of pixels $N$, the image size and the frame rate (FPS). The pixel size and resolution may be adjusted by changing the spot size of the pump laser in the diamond and the thickness of the NV-enriched layer.

Fluorescence from the NV centers is collected through the same objective and spectrally filtered using a longpass filter (FEL600, Thorlabs) to remove residual \SI{532}{\nano\meter} pump light. Approximately \SI{50}{\micro\watt} of NV fluorescence is detected using a variable-gain linear photodetector (OE-300-SI, FEMTO). For the data presented in this article, the detector operated at a transimpedance gain of \(10^5~V/A \), yielding a detection bandwidth of \SI{3.5}{\mega\hertz}.

A beam splitter (BS) positioned after the AOD diverts a small portion of the green excitation beam to a second photodetector identical to the one used for fluorescence detection. The outputs of the detectors are digitized independently using a Spectrum Instruments DAQ card with a sampling rate of \SI{20}{\mega\sample\per\second} and \num{16} bit resolution. The DAQ and AWG are synchronized to ensure precise temporal alignment between the scan and the detected signal, allowing us to implement a common-mode rejection (CMR) scheme in software for the suppression of correlated technical laser noise. Microwave excitation is provided via a signal generator (SG394, Stanford Research Instruments) and an amplifier (ZHL-16W-43-S+, Mini-Circuits; \SI{16}{\watt} output), driving the resonator beneath the diamond for NV center electron spin manipulation. A pair of neodymium permanent magnets is positioned across the diamond and used to apply a static magnetic field with a strength $\sim$\SI{3}{\milli\tesla} aligned along a [110] crystallographic direction. This field configuration produces two pairs of degenerate NV spin-resonances and ensures consistent spin projection across the ensemble, enhancing both readout contrast and sensitivity to magnetic fields compared to alignment along [111].

\subsection{Scanning and Data Acquisition}
Scanning of the optical excitation beam with the AOD in the NV-enriched layer follows a serpentine (bidirectional) raster pattern, in which the beam moves across each row in one direction and then reverses direction for the subsequent row. With the total number of pixels per frame $N$ and the pixel dwell time $\tau_d$, the frame rate (FPS) is simply given by the relation  $\text{FPS} = (N \tau_d)^{-1}$. We tested values of \(\tau_d\) as low as \SI{0.4}{\micro\second}, limited by the beam diameter in the AOD crystal and the resulting rise time. For images with \(N = 11\times 11\) pixels, this results in frame rates of up to \SI{20}{\kilo\hertz}.

We calibrated and dynamically adjusted the AOD drive amplitude for each pixel to compensate for spatial variations in laser intensity and NV density. This optimized approach ensures a consistent
level of detected fluorescence across the entire image.

\subsection{Magnetic Field Imaging}

To demonstrate the capability of imaging spatiotemporal magnetic fields, we positioned a \SI{20}{\micro\meter}-diameter copper wire directly on the diamond surface and applied a signal generated with a DAC (National Instruments) to induce a spatially and temporally varying magnetic field.

The generated square signal had a total duration of \SI{50}{\milli\second}, an amplitude of \SI{0.01}{\volt}, and inverted polarity halfway through the pulse. The electrical load was measured to be \SI{5}{\ohm}, corresponding to a drive current of approximately \SI{2}{\milli\ampere}. The pulses were synchronized with the magnetic field imaging using a digital marker output from the AWG. The magnetic field induced by the current in the wire was mapped in real time using our scanning NV magnetometry technique. For this experiment, a dwell time of $\tau_d = \SI{1.2}{\micro\second}$ was used with a \(21\times21\) pixel grid, resulting in a frame rate of \SI{2}{\kilo\hertz}. The scanned region covered \(\sim\)\SI{170}{\micro\meter}\(\times\)\SI{170}{\micro\meter}, with a Gaussian beam spot size of \SI{10}{\micro\meter} and a center-to-center pixel spacing of \SI{8}{\micro\meter}, yielding an effective overlap of approximately \SI{15}{\percent} between adjacent pixels. This degree of overlap provided sub-beam sampling and was consistent with the Nyquist criterion for raster-scanned imaging, preventing undersampling and reconstruction artifacts~\cite{pawley2006handbook}.

We recorded \SI{0.1}{\second}-long time-traces for each pixel and averaged them to evaluate noise scaling. The time-traces were additionally filtered using narrowband notch filters to suppress correlated magnetic and electrical pickup at discrete frequencies (e.g., the mains frequency and harmonics).

The optimal microwave (MW) frequency for sensing was selected by computing the spatial average of the ODMR spectra across all pixels and identifying the frequency with the steepest slope. This frequency was then used for subsequent sensing measurements. To suppress pixel-to-pixel amplitude variations arising from spatial differences in optical collection efficiency, NV density, and local microwave detuning, each time-trace was normalized to the response generated by a weak externally applied sinusoidal magnetic field at \SI{727}{\hertz}. This calibration field was produced by a pair of coils positioned sufficiently far from the sensing region so that the resulting magnetic field could be considered homogeneous across the entire scanned area. The procedure equalized the relative signal scaling between pixels but did not alter the intrinsic magnetic sensitivity, which remained limited by the local noise level at each pixel.

\section{\label{sec:level3}Results}

\subsection{Quasi-Continuous Wave ODMR}

\begin{figure*}[t]  
    \centering
    \includegraphics[width=0.9\textwidth]{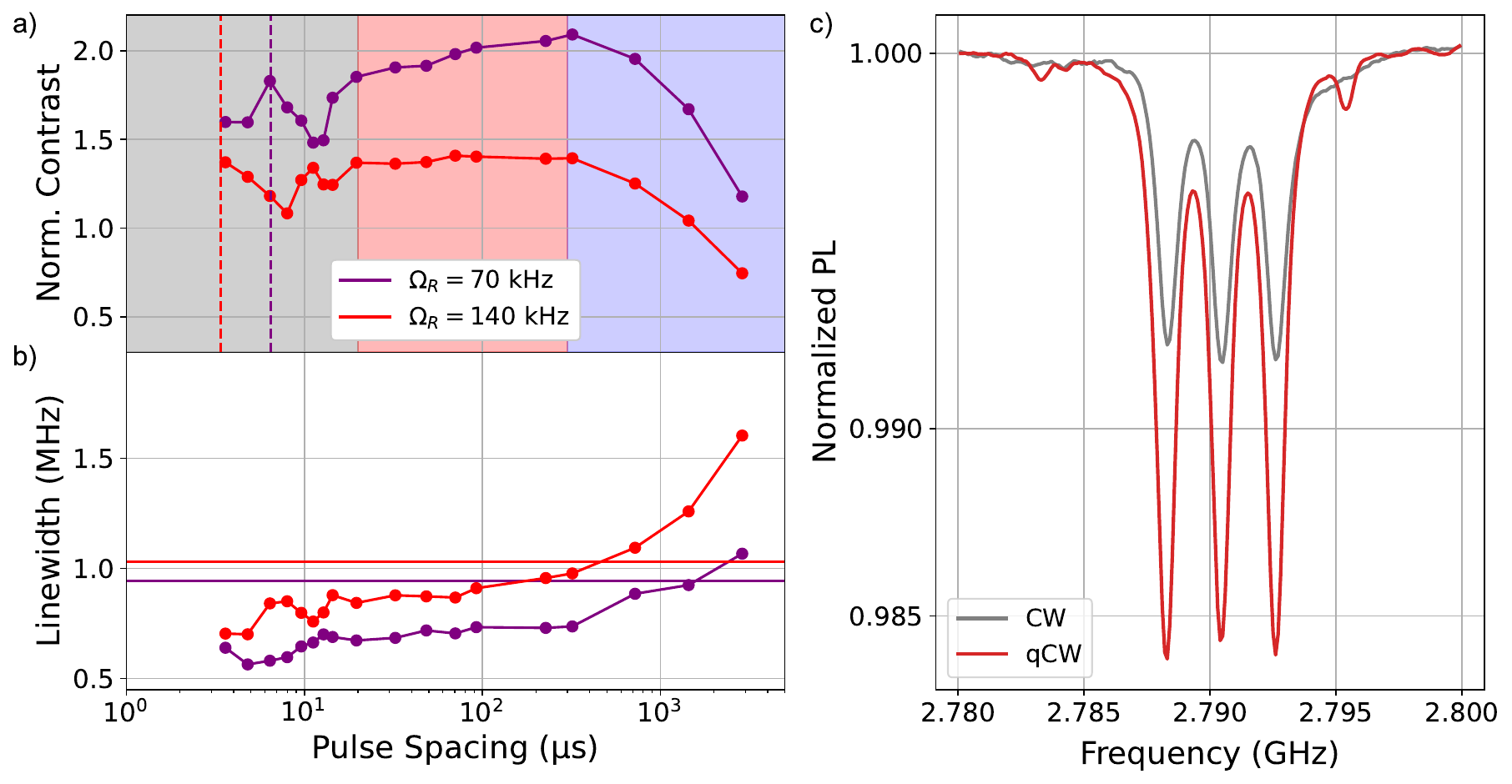} 
\caption{
Quasi-continuous-wave (qCW) ODMR characterization of a single pixel with the magnetic field aligned along the [110] crystal direction. 
(a) Normalized contrast and (b) linewidth, shown as a function of laser pulse spacing for two Rabi frequencies $\Omega_R$. Pulse spacing defines the time between successive \SI{0.4}{\micro\second} laser pulses, while the microwave field is applied continuously. 
In (a), contrast is normalized to the CW-ODMR contrast at the corresponding $\Omega_R$.
Vertical dashed lines indicate the expected $\pi$-pulse condition. 
Shaded regions represent the approximate regimes of coherent Rabi oscillations (gray), dephasing-dominated evolution (red), and relaxation-limited decay ($T_1$ regime, blue). 
In (b), solid horizontal lines indicate the corresponding linewidths obtained in CW for each Rabi frequency.
(c) ODMR spectra of the $m_s = 0$ to $-1$ transition acquired under a Rabi drive of \(\Omega_R = 70~\text{kHz}\) for a single pixel. The CW (purple) and qCW (red) spectra are shown; the qCW ODMR is measured with a pulse spacing of \SI{100}{\micro\second}, revealing enhanced contrast compared to CW.
}
    \label{fig:figure2}  
\end{figure*}

Our magnetic field imaging protocol combines continuous wave microwave excitation with spatially and temporally localized optical excitation of NV centers. The temporal spacing between successive optical excitation pulses is set by the scan time per frame, while the pulse duration is fixed by the pixel dwell time. This excitation scheme shares conceptual similarities with the pulsed-ODMR method introduced by Dréau et al.~\cite{dreau2011avoiding}, where interleaved microwave and optical readout pulses suppress power broadening and recover the inhomogeneous linewidth. In contrast, our microwaves are left on continuously, and the optical pulses arise naturally from the beam-scanning dwell time rather than from explicit pulse gating, producing a quasi-CW regime in which coherent evolution and relaxation occur between successive optical resets. To characterize the performance of the protocol, we extract the ODMR spectrum at one pixel and determine the magnetic-field sensitivity from the ODMR contrast and linewidth as a function of the scan time. It is important to note that in our scanning configuration with a fixed pixel dwell time $\tau_d$, the total number of pixels per frame directly determines the pulse spacing time. Thus, for constant $\tau_d$, increasing the image size or spatial resolution (number of pixels) inherently increases the pulse spacing time for each pixel, affecting the ODMR readout properties. The measured ODMR contrast and linewidth are summarized in Fig.~\ref{fig:figure2}a and \ref{fig:figure2}b, respectively. 

With zero pulse spacing, both linewidth and contrast are determined by the steady state conditions in the continuous wave (CW) regime~\cite{webb2020optimization, ahmadi2017pump}. For a fixed dwell time of \SI{0.4}{\micro \second} and with increasing pulse spacing, we observe an overall increase of contrast as compared to the CW regime. A maximum occurs for a pulse spacing corresponding to a nominal $\pi$-pulse of the spin Rabi frequency $\Omega_R$. We tested $\Omega_R$ values of \SI{70}{\kilo\hertz} and \SI{140}{\kilo\hertz}, respectively, while the optical excitation rate is $\Gamma_p \approx \SI{1}{\mega\hertz}$ under our illumination conditions.
The optical pulse effectively polarizes the NV centers into the bright $m_s = 0$ spin-state, while between two optical pulses the spin population is coherently driven by the continuously applied microwave field. However, the maximum contrast does not occur at the $\pi$-point of the corresponding Rabi frequency, but instead appears at a significantly longer pulse spacing time around \SI{100}{\micro\second} in our case. We associate the resulting delayed contrast peak to inhomogeneous broadening of the ensemble NV center spin transitions larger than the applied $\Omega_R$, reducing the efficiency of coherent population transfer and shifting the timing of maximum readout contrast. This interpretation is supported by our numerical simulations (see Supplementary Materials), which reproduce the delayed contrast maximum and show that microwave detuning across the ensemble modifies the effective rotation rate and damps the oscillation envelope.

As a function of optical pulse spacing, the linewidth (full width at half maximum) initially becomes narrower than in the CW limit and then broadens with increasing pulse spacing. This observation reflects a transition from coherent to relaxation-dominated dynamics. 
Figure~\ref{fig:figure2}c compares the ODMR spectrum of a single pixel acquired under CW and qCW conditions with a Rabi frequency of \(\Omega_R = \SI{70}{\kilo\hertz}\). The qCW spectrum, taken at a pulse spacing of \SI{100}{\micro\second}, exhibits enhanced contrast consistent with the dynamics described above.

We use a 5-level model~\cite{ahmadi2017pump, robledo2011spin} to simulate the evolution and readout of the spin states. This model qualitatively captures the key features observed in our experiments, including the position of the $\pi$-pulse peak and the subsequent contrast maximum (see Supplementary). This agreement supports the interpretation that compared to the CW-regime, contrast enhancement arises from coherent spin dynamics related to the qCW regime applied. At longer pulse spacings, the contrast decreases as longitudinal spin relaxation reduces net polarization before the next readout, indicating a spin-lifetime $T_1$-limited performance.

\begin{figure}[t]  
    \centering
    \includegraphics[width=0.6\linewidth]{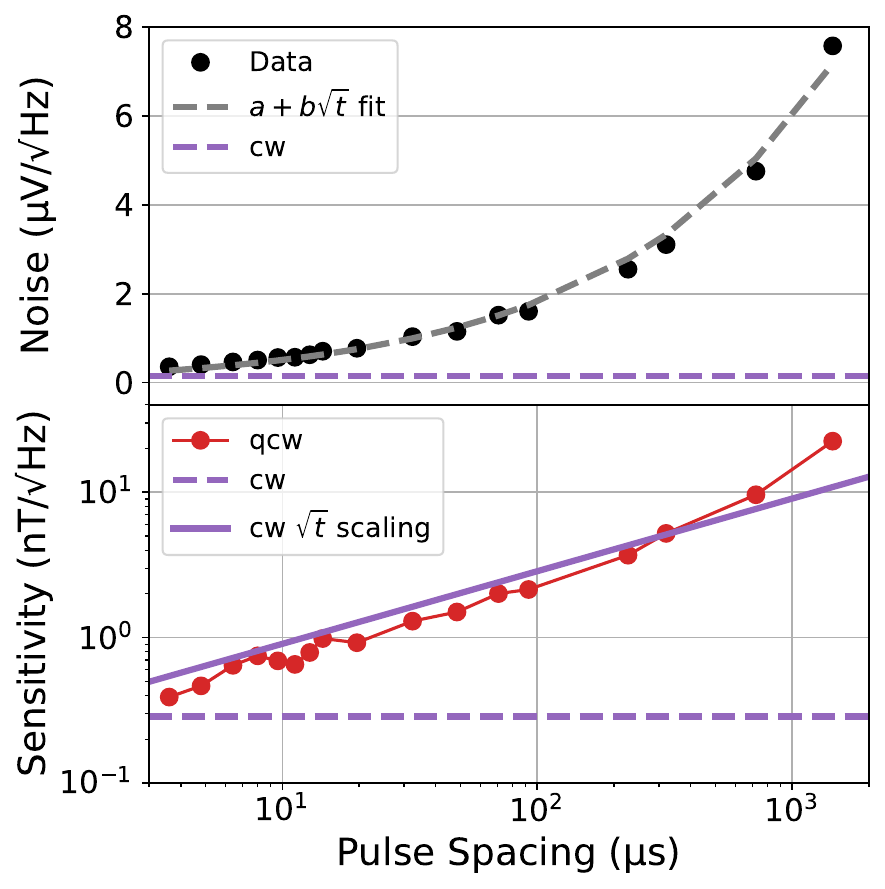} 
    \caption{
Top: Noise floor levels (in \(\mu\text{V}/\sqrt{\text{Hz}}\)) for a single pixel as a function of pulse spacing, with total measurement time fixed at 1\,s. As pulse spacing increases, fewer photons are collected, leading to a noise scaling consistent with \(\sqrt{t}\). The dashed line shows the CW noise floor. 
Bottom: Sensitivity versus pulse spacing. The dashed purple line indicates the CW sensitivity baseline. The solid purple line shows the expected CW \(\sqrt{t}\) scaling assuming constant contrast and linewidth. The red curve represents the actual sensitivity in the qCW regime, which initially improves beyond the expected CW scaling before degrading at longer spacings due to increasing noise and reduced contrast.
}
\label{fig:figure3}  
\end{figure}

To quantify the magnetic sensing performance, we extract noise floor levels from \SI{1}{\second} time-traces as a function of pulse spacing (Fig.~\ref{fig:figure3}a). The observed \(\sqrt{t}\) scaling as a function of pulse spacing reflects the reduction in collected photons at longer spacings over a fixed total measurement time. The CW limit with \SI{0.16}{\micro\volt\per\sqrt\hertz} provides a baseline for comparison.

The corresponding sensitivity trends are summarized in Fig.~\ref{fig:figure3}b. The solid purple curve indicates the scaling with fixed contrast and linewidth that would have been obtained from CW reference measurements and accounting for reduction in collected photons with \(\sqrt{t}\)-scaling, while the dashed line marks the \SI{200}{\pico\tesla\per\sqrt\hertz} sensitivity of the single pixel measured in the CW regime. For short pulse spacings, less than $\sim$\SI{300}{\micro \second}, our qCW data outperforms the extrapolated CW behavior (constant linewidth and contrast), benefiting from enhanced contrast, but ultimately degrades at longer spacings due to reduced photon counts and decreased spin-polarization. These results highlight the trade-offs between coherent contrast enhancement and relaxation-induced sensitivity loss in the qCW-regime.

\subsection{Imaging of Magnetic Fields from a Microwire}

\begin{figure*}
    \centering
    \includegraphics[width=0.9\textwidth]{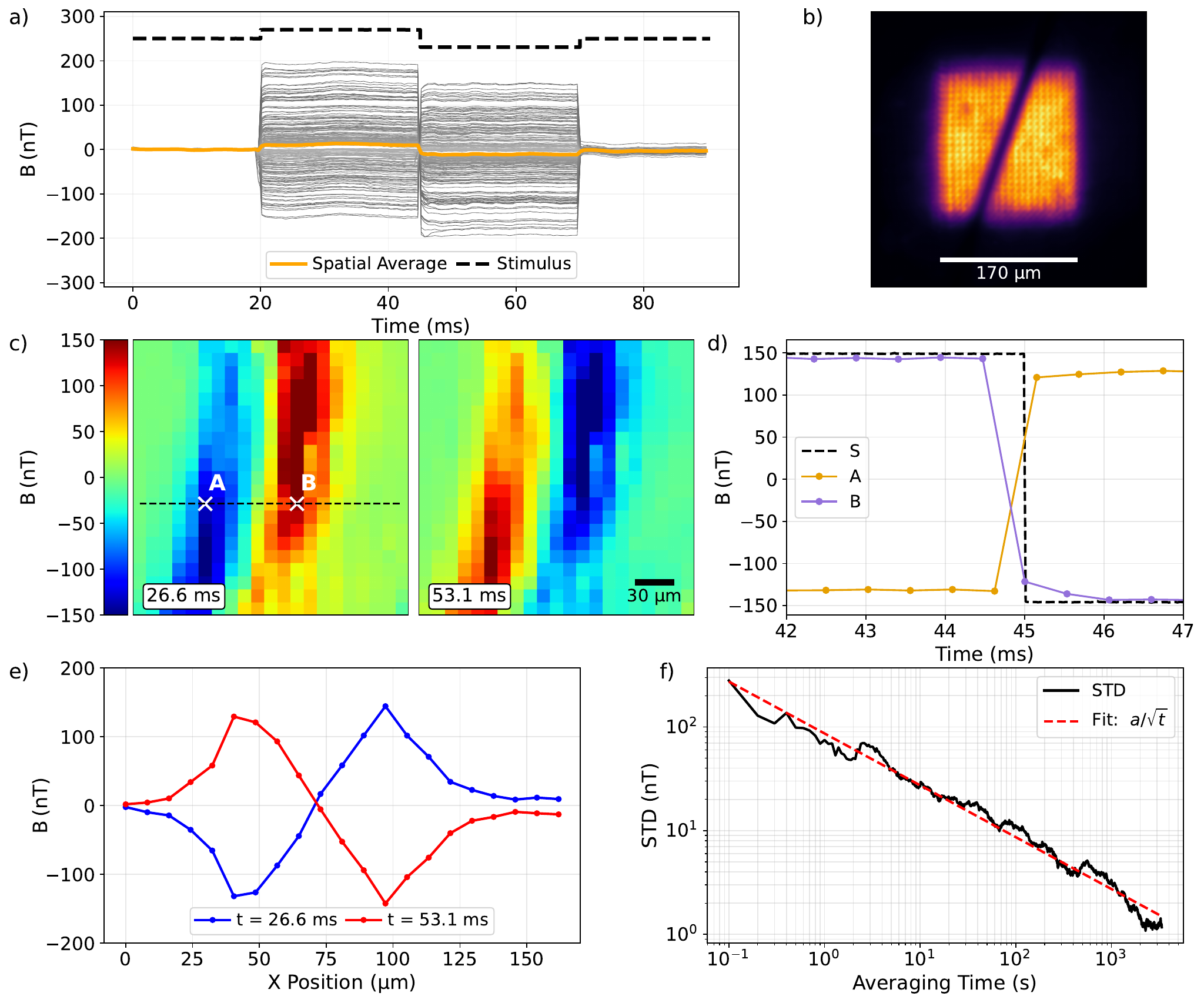}
    \caption{ Wide-field NV magnetometry of a current-carrying microwire.
a) Time-traces from all 441 pixels in the $21\times 21$ scan (gray). The orange curve represents the spatial average of all traces, and the black dashed curve shows the recorded biphasic current stimulus.
b) Microscope image showing the microwire relative to the NV sensing region. 
The oblique dark line indicates the wire position; scale bar: \SI{170}{\micro\meter}.
c) Magnetic field maps at two selected times (26.6 ms and 53.1 ms), corresponding to the positive 
and negative phases of the applied stimulus. Pixels A and B (white X marks) lie on a horizontal cut 
(dashed line) used for subsequent analysis.
d) Time-traces of pixels A (gold) and B (purple), overlaid with the applied stimulus (black). 
The two traces show opposite magnetic polarity due to their positions on opposite sides of the wire.
e) Horizontal line profiles across the sensing plane at the same two timestamps as in (c), 
revealing the characteristic dipole-like magnetic field generated by the microwire.
f) Noise performance of the sensor array. The standard deviation of successive averages follows 
the expected $1/\sqrt{t}$ scaling (red dashed line), demonstrating stable temporal averaging over 
three decades in time.
}
    \label{fig:figure4}
\end{figure*}

Having established the behavior and performance of the qCW-ODMR protocol at the single-pixel level, we now demonstrate its applicability to wide-field magnetic imaging by mapping the spatiotemporal magnetic field generated by a current-carrying microwire. Using the wire-stimulation setup described above, we recorded time-resolved maps of the magnetic-field projection along the effective sensing axis, \(B_{[110]}\), during the application of square pulses with a phase inversion in the center to a \SI{20}{\micro\meter} copper wire. The measurements were performed using our AOD-based scanning technique, giving a mean per-pixel sensitivity of \SI{10}{\nano\tesla\per\sqrt\hertz} in the \SIrange{100}{1000}{\hertz} band.

Figure~\ref{fig:figure4}a shows the reconstructed time-traces from all \num{441} pixels during two periods of the biphasic stimulus, averaged over \SI{20}{\minute}. Individual pixels exhibit clear magnetic-field responses, whereas the spatial average (orange) remains comparatively small—a direct consequence of the strongly varying field produced by the wire, underscoring the necessity of spatially resolved detection. The microscope image in Fig.~\ref{fig:figure4}b is taken from above the diamond and shows the relative position of the wire across the \(\sim170\times\SI{170}{\micro\meter\squared}\) scanned area.

Magnetic-field maps at two representative times in the current cycle are shown in Fig.~\ref{fig:figure4}c. The maps display the expected dipolar structure and the polarity reversal between the positive and negative phases of the pulse. Two representative pixels, A and B, located on opposite sides of the wire, are highlighted. Their time-traces (Fig.~\ref{fig:figure4}d) exhibit opposite magnetic polarity, and both faithfully follow the rapid transitions of the \SI{10}{\kilo\hertz}-sampled stimulus, demonstrating preserved temporal response during abrupt switching of the phase in the applied signal.

To quantify the spatial field structure, horizontal linecuts extracted at the two times indicated in Fig.~\ref{fig:figure4}c are shown in Fig.~\ref{fig:figure4}e. The profiles exhibit sharp extrema at the wire location, consistent with the steep gradient predicted by the law of Biot--Savart for a thin conductor. The narrowness of these features confirms that the system’s spatial resolution—set by the \SI{10}{\micro\meter} optical spot and \SI{8}{\micro\meter} pixel spacing—is sufficient to clearly resolve the phase in the induced magnetic field distribution.

Finally, Fig.~\ref{fig:figure4}f shows the noise scaling obtained from repeated measurements. The standard deviation decreases as \(1/\sqrt{\tau}\), indicating stable averaging behavior and the absence of long-term technical drifts over the full integration range.

\section{\label{sec:level4}Discussion}
We have demonstrated a scanning NV magnetometry technique based on acousto-optic beam deflection and quasi-continuous-wave ODMR (qCW-ODMR), enabling high-speed magnetic imaging with enhanced sensitivity and dynamic range. Software-based common-mode rejection suppresses laser-intensity noise, improving measurement stability.

The qCW protocol combines continuous microwave excitation with spatially localized optical readout and repolarization. This approach removes the need for precise microwave timing control and avoids the use of gated microwave pulses or arbitrary waveform generators. Instead, the inter-pulse spacing—set directly by the scan configuration—governs the spin dynamics. As shown experimentally and through numerical modeling, the system transitions from coherent microwave-driven evolution to a \(T_1\)-dominated regime as the pulse spacing increases.

Because the number of scan points determines the inter-pulse spacing, the imaging approach introduces an inherent trade-off between spatial resolution, image size, and sensitivity. Increasing the pixel number reduces the frame rate and increases the effective pulse spacing, which modifies the ODMR contrast and linewidth (Fig.~\ref{fig:figure2}) and therefore the sensitivity. Additional spatial inhomogeneities arise from variations in strain, microwave amplitude, optical intensity, and detuning, since all NVs are driven by a single global microwave frequency. We mitigate these effects by choosing the microwave drive frequency based on spatially averaged ODMR spectra and by normalizing each pixel to a homogeneous calibration field generated by external coils.

As a functional demonstration of the system’s capabilities, we imaged the magnetic field induced from the current in a microwire. The measured spatiotemporal maps reproduced the expected polarity and evolution of the field, while the linecuts revealed sharp spatial features consistent with the predicted gradient near the wire. The pixel-level traces followed the rapid current reversals, confirming preserved temporal response, and the averaging analysis showed stable noise reduction. Together, these results validate the practical imaging performance of the platform.

\begin{figure} 
    \centering
    \includegraphics[width=0.7\linewidth]{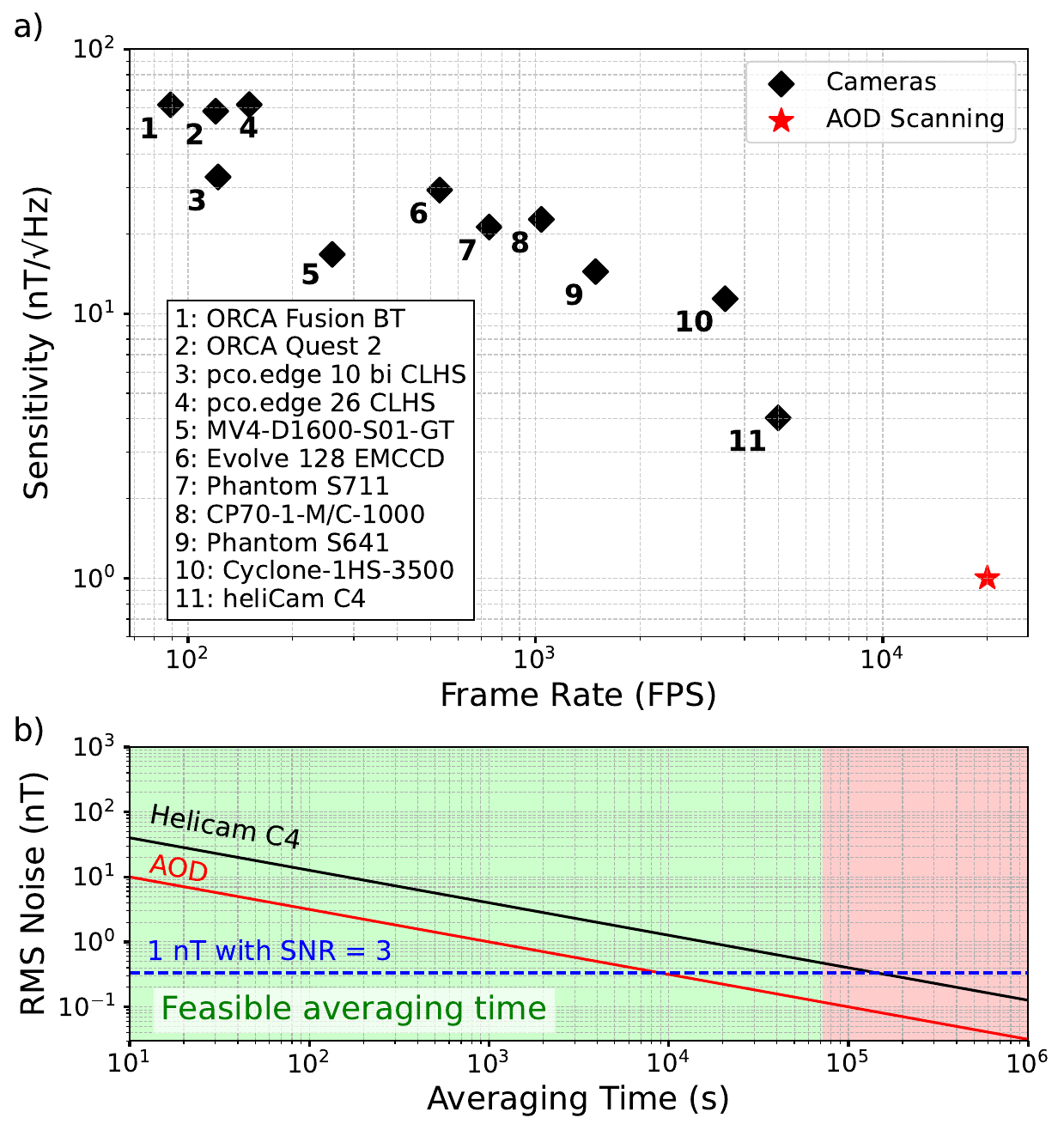} 
  \caption{
        \textbf{(a)} Magnetic sensitivity versus frame rate for conventional camera-based systems and our AOD scanning approach. Black diamonds represent state-of-the-art scientific and high-speed cameras, with sensitivity normalized to the effective area of a single AOD pixel (\SI{30}{\micro\meter}), treating camera macropixels as matched sensing volumes. Labels correspond to camera models listed in the inset. The red star marks the performance of our AOD method, which achieves superior sensitivity at higher frame rates. 
        \textbf{(b)} Projected root-mean-square (RMS) magnetic noise versus averaging time for our AOD system and a representative high-end camera (heliCam C4). The green region marks averaging times up to \SI{24}{\hour}, corresponding to the practical upper limit for continuous measurement in biological samples. The red region indicates integration times beyond this limit. The dashed line marks a \SI{1}{\nano\tesla} detection threshold at signal-to-noise ratio (SNR) = 3. The AOD approach achieves lower noise and faster averaging compared to camera-based systems.
    }
    \label{fig:figure5} 
\end{figure}
To benchmark the technique, we compared the sensitivity and frame rate of the AOD-scanning system with those of state-of-the-art cameras commonly used in NV magnetometry (Fig.~\ref{fig:figure5}a). Sensitivities were normalized by pixel area and field of view to enable a fair comparison across detectors. Our system achieves substantially better sensitivity at significantly higher frame rates due to its high dynamic range, fast beam steering, and optimized readout strategy. A detailed description of the calculation procedure is provided in the Supplementary Materials. Figure~\ref{fig:figure5}b illustrates the practical implications: to reach an SNR of 3 for a \SI{1}{\nano\tesla} signal with \SI{1}{\hertz} averaging rate, the AOD system requires approximately \SI{2.5}{\hour} of averaging, whereas a high-end lock-in camera (heliCam C4) would require more than \SI{40}{\hour} under ideal conditions. Such extended acquisition times exceed typical biological viability windows of \(\sim\) \SI{24}{\hour}, underscoring the advantage of the AOD approach for detecting weak, time-varying magnetic fields in biological preparations.

Looking ahead, several improvements could further enhance the performance of the platform. Sensitivity may be increased by implementing lock-in detection schemes based on modulation of the laser and the microwave drive, which would suppress low-frequency noise and improve long-term stability. Double-quantum (DQ) readout, enabled by a dual-frequency microwave drive, would offer an additional gain in magnetic sensitivity and reduced common-mode noise. Further gains could be achieved by improving optical collection efficiency, optimizing beam shaping, or adopting adaptive scan strategies that allocate longer dwell times to regions of interest. Together, these developments would push the technique toward higher sensitivity, larger fields of view and higher frame rates.

\section{\label{sec:level5}Conclusion}

Our work introduces a scanning NV magnetometry platform that combines fast AOD beam steering, high-speed photodetection, and a quasi-continuous-wave ODMR (qCW-ODMR) protocol, enabling magnetic imaging at frame rates above \SI{10}{\kilo\hertz} with nT-level sensitivity. By scanning a focused laser beam across an ensemble of NV centers under continuous microwave excitation, each pixel experiences a brief, localized optical reset. This configuration accesses a tunable spin regime in which the contrast and sensitivity are modulated by the pulse spacing between successive excitations.

We show that the qCW-regime enables flexible optimization of sensitivity and bandwidth, and supports frame rates exceeding \SI{10}{\kilo\hertz} across a significant number of pixels. The microwire measurement provides a proof-of-principle demonstration of the platform, showing that the sensor preserves both spatial precision and kHz-band temporal response under realistic experimental conditions. With a mean pixel sensitivity of \SI{10}{\nano\tesla\per\sqrt\hertz} in the 100–1000~Hz band, the system resolves weak, rapidly evolving magnetic fields on micrometer length scales.

Compared to conventional wide-field camera-based approaches, our system achieves superior sensitivity at significantly higher frame rates, as we confirm by a fair performance comparison across multiple commercial cameras. 

Overall, our platform provides a versatile and scalable framework for high-speed magnetic imaging in environments where weak, time-varying signals must be captured with both spatial and temporal precision, ultimately limited by diffraction and the NV spin dynamics. The ability to flexibly tune resolution and sensitivity makes our magnetic field imaging approach particularly well suited for applications in biomagnetic imaging.

\begin{acknowledgments}
\noindent We gratefully acknowledge support from the Novo Nordisk Foundation (Biomag NNF - 21OC0066526, CBQS NNF24SA0088433, MagGen NNF23OC0084195) and the Danish National Research Foundation (DNRF) through the center for Macroscopic Quantum States (bigQ, Grant No. DNRF0142).

\end{acknowledgments}

\bibliographystyle{apsrev4-2}
\bibliography{references}

\end{document}